\documentclass[sigconf]{acmart}

\AtBeginDocument{%
  }

\copyrightyear{2025}
\acmYear{2025}
\setcopyright{cc}
\setcctype{by}
\acmConference[CHI '25]{CHI Conference on Human Factors in Computing Systems}{April 26-May 1, 2025}{Yokohama, Japan}
\acmBooktitle{CHI Conference on Human Factors in Computing Systems (CHI '25), April 26-May 1, 2025, Yokohama, Japan}\acmDOI{10.1145/3706598.3713664}
\acmISBN{979-8-4007-1394-1/25/04}

\usepackage{amsmath}
\usepackage{enumitem}
\usepackage{xcolor}
\usepackage{listings}

\usepackage{tikz}
\usepackage{soul}
\usepackage{tikz}
\usetikzlibrary{calc}
\usetikzlibrary{shapes.misc}
\usetikzlibrary{fit}

\makeatletter
\newbox\mybox
\newcount\test
\newcommand{\defhighlighter}[3][]{%
    \tikzset{every highlighter/.style={color=#2, fill opacity=#3, #1}}%
}

\defhighlighter[font=\ttfamily]{backcolour}{1}

\newcommand{\highlight@DoHighlight}[1]{
    \node[outer sep = 0pt, inner xsep = 3pt, inner ysep = 0pt,
        fit=(begin highlight) (end highlight),
        every highlighter, this highlighter,
        rectangle, rounded corners=1mm,#1,
        fill
    ]{} ;
}

\newcommand{\highlight@BeginHighlight}[1]{
  \coordinate[yshift=-\dp\mybox-1pt,#1] (begin highlight) at (0,0) ;
}

\newcommand{\highlight@EndHighlight}[1]{
  \coordinate[yshift=\ht\mybox+1pt,#1 ] (end highlight) at (0,0) ;
}

\newdimen\highlight@previous
\newdimen\highlight@current

\DeclareRobustCommand*\highlight[2][]{%
  \nobreak\hspace{.15em plus .08333em}%
  \setbox\mybox\hbox{#2}%
  \tikzset{this highlighter/.style={#1}}%
  \SOUL@setup%
  \gdef\ht@possiblenoleftround{}%
  \def\SOUL@preamble{%
    \begin{tikzpicture}[overlay, remember picture, blend mode=darken]
      \highlight@BeginHighlight{xshift=1pt}
      \highlight@EndHighlight{xshift=-1pt}
    \end{tikzpicture}%
  }%
  \def\SOUL@postamble{%
    \begin{tikzpicture}[overlay, remember picture, blend mode=darken]
      \highlight@EndHighlight{xshift=-1pt}
      \expandafter\highlight@DoHighlight\expandafter{\ht@possiblenoleftround}
    \end{tikzpicture}%
  }%
  \def\SOUL@everyhyphen{%
    \discretionary{%
      \SOUL@setkern\SOUL@hyphkern%
      \SOUL@sethyphenchar%
      \tikz[overlay, remember picture, blend mode=darken] \highlight@EndHighlight{} ;%
    }{%
    }{%
      \SOUL@setkern\SOUL@charkern%
    }%
  }%
  \def\SOUL@everyexhyphen##1{%
    \SOUL@setkern\SOUL@hyphkern%
    \hbox{##1}%
    \discretionary{%
      \tikz[overlay, remember picture, blend mode=darken] \highlight@EndHighlight{} ;%
    }{%
    }{%
      \SOUL@setkern\SOUL@charkern%
    }%
  }%
  \def\SOUL@everysyllable{%
    \begin{tikzpicture}[overlay, remember picture, blend mode=darken]
      \path let \p0 = (begin highlight), \p1 = (0,0) in \pgfextra
        \global\highlight@previous=\y0
        \global\highlight@current =\y1
      \endpgfextra (0,0) ;
      \ifdim\highlight@current < \highlight@previous
        \expandafter\highlight@DoHighlight\expandafter{\ht@possiblenoleftround,rounded rectangle right arc=none}
        \gdef\ht@possiblenoleftround{rounded rectangle left arc=none}
        \highlight@BeginHighlight{}
      \fi
    \end{tikzpicture}%
    \ttfamily \the\SOUL@syllable%
    \tikz[overlay, remember picture, blend mode=darken] \highlight@EndHighlight{} ;%
  }%
  \SOUL@{#2}%
  \nobreak\hspace{.15em plus .08333em}%
}

\newcommand{\tqlinline}[1]{\highlight{\texttt{#1}}}

\definecolor{codegreen}{rgb}{0,0.6,0}
\definecolor{codegray}{rgb}{0.5,0.5,0.5}
\definecolor{codepurple}{rgb}{0.58,0,0.82}
\definecolor{backcolour}{rgb}{0.937,0.953,0.9725}
\definecolor{framecolour}{rgb}{0.282,0.327,0.392}

\lstdefinestyle{tempoql}{
    backgroundcolor=\color{backcolour},   
    commentstyle=\color{codegreen},
    keywordstyle=\color{magenta},
    numberstyle=\tiny\color{codegray},
    stringstyle=\color{codepurple},
    basicstyle=\ttfamily\footnotesize,
    breakatwhitespace=false,         
    breaklines=true,                 
    captionpos=b,         
    keepspaces=true,                 
    showspaces=false,                
    showstringspaces=false,
    showtabs=false,                  
    tabsize=2,
    frame=tb,
    rulecolor=\color{framecolour},
    framextopmargin=2ex,
    framexbottommargin=2ex,
    aboveskip=8pt,
    belowskip=8pt
}
\begin{document}

\title{Tempo: Helping Data Scientists and Domain Experts Collaboratively Specify Predictive Modeling Tasks}

\author{Venkatesh Sivaraman}
\orcid{0000-0002-6965-3961}
\affiliation{%
  \institution{Carnegie Mellon University}
  \city{Pittsburgh}
  \state{Pennsylvania}
  \country{USA}
}

\author{Anika Vaishampayan}
\affiliation{%
  \institution{Carnegie Mellon University}
  \city{Pittsburgh}
  \state{Pennsylvania}
  \country{USA}
}

\author{Xiaotong Li}
\affiliation{%
  \institution{University of Pittsburgh}
  \city{Pittsburgh}
  \state{Pennsylvania}
  \country{USA}
}

\author{Brian R Buck}
\affiliation{%
  \institution{University of Pittsburgh}
  \city{Pittsburgh}
  \state{Pennsylvania}
  \country{USA}
}

\author{Ziyong Ma}
\affiliation{%
  \institution{Carnegie Mellon University}
  \city{Pittsburgh}
  \state{Pennsylvania}
  \country{USA}
}

\author{Richard D Boyce}
\affiliation{%
  \institution{University of Pittsburgh}
  \city{Pittsburgh}
  \state{Pennsylvania}
  \country{USA}
}

\author{Adam Perer}
\affiliation{%
  \institution{Carnegie Mellon University}
  \city{Pittsburgh}
  \state{Pennsylvania}
  \country{USA}
}

\renewcommand{\shortauthors}{Sivaraman et al.}

\begin{abstract}
Temporal predictive models have the potential to improve decisions in health care, public services, and other domains, yet they often fail to effectively support decision-makers. 
Prior literature shows that many misalignments between model behavior and decision-makers' expectations stem from issues of model specification, namely how, when, and for whom predictions are made. 
However, model specifications for predictive tasks are highly technical and difficult for non-data-scientist stakeholders to interpret and critique. 
To address this challenge we developed Tempo, an interactive system that helps data scientists and domain experts collaboratively iterate on model specifications. 
Using Tempo's simple yet precise temporal query language, data scientists can quickly prototype specifications with greater transparency about pre-processing choices. 
Moreover, domain experts can assess performance within data subgroups to validate that models behave as expected. 
Through three case studies, we demonstrate how Tempo helps multidisciplinary teams quickly prune infeasible specifications and identify more promising directions to explore.
\end{abstract}


\begin{teaserfigure}
  \centering
  \includegraphics[width=0.72\textwidth, alt={Screenshot of Tempo applied to predicting readmission in patients in home health care, with arrows from data scientists to the Models sidebar and from domain experts to the Subgroups view. The top subgroup returned for positive predictions by the 30-day readmission model is patients aged 65-75, with 5-10 days in the hospital in the past 6 months. These patients are predicted to be readmitted 91\% of the time compared to 47\% on average.}]{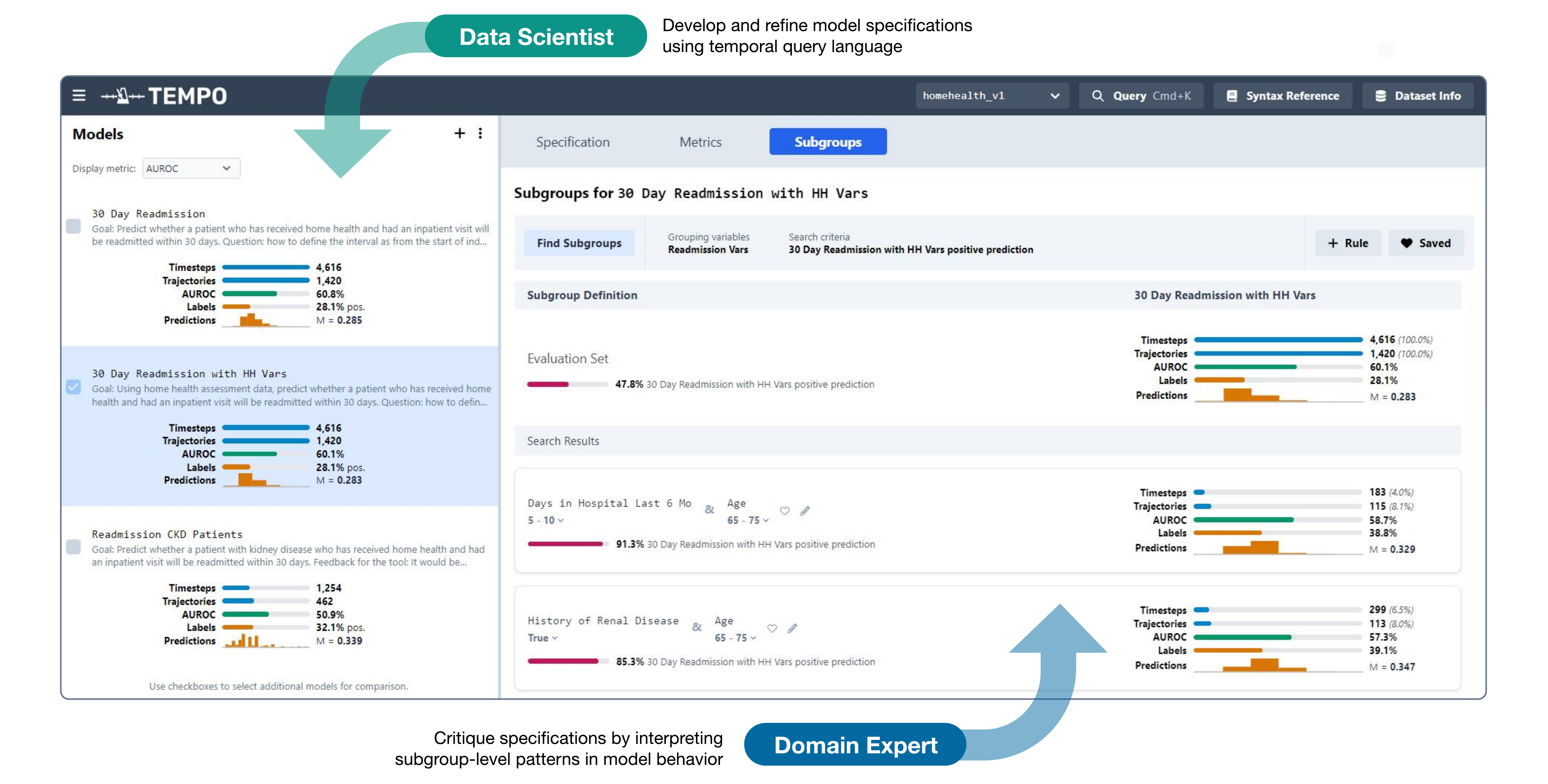} 
  \caption{Tempo helps data scientists define and evaluate specifications of predictive modeling tasks by bringing non-technical domain experts into the loop early in development. Here, we worked with pharmacy experts to predict hospital readmission for patients receiving home health care (see Sec. \ref{sec:case-study-home-health}). Data scientists on the team developed predictive model specifications using Tempo's query language, enabling rapid experimentation with different patient cohorts and target variables. Pharmacy experts could then critique the models' design and suggest improvements by examining subgroups most associated with readmission.}
  \label{fig:teaser}
\end{teaserfigure}


\maketitle

\section{Introduction} \label{sec:introduction}

Predictive AI models aim to enhance human decision-making by inferring individuals' future outcomes, relating their past observations against vast amounts of historical data.
In expert communities, ranging from medical and public health professionals to public services agencies to business analysts, predictive capabilities are increasingly expected to streamline and standardize decision processes.
However, many decision support tools (DSTs) built on predictive models fail to effectively support decision-makers in practice. 
While in some cases these failures can be attributed to the nature of the AI's deployment, such as incompatibility with existing workflows or poor usability~\cite{dziorny_clinical_2022,Khairat2018,Wang2021}, negative perceptions of predictive models often stem from more fundamental concerns about their behavior.
In particular, issues often arise when existing decision processes conflict with the \textit{model specification}, or the precise task that the predictive model is trained to perform. 
Model specifications determine \textit{what}, \textit{how}, and \textit{for whom} predictions will be made~\cite{sherman_leveraging_2018,tal_target_2023}, all of which can impact how decision-makers perceive the resulting models to behave. 
In high-stakes fields such as medical treatment~\cite{sherman_leveraging_2018,Sivaraman2023}, child welfare~\cite{Kawakami2022partnerships}, and housing allocation~\cite{Kuo2023,cheng_2024_aha}, misaligned model specifications can lead experts to avoid using the models out of concern that they will introduce or perpetuate systemic decision-making errors.
Therefore, helping data scientists and domain experts collaboratively define the right predictive modeling task is key to developing DSTs that decision-makers find relevant and useful.

To demonstrate how these obstacles might play out in a practical scenario (inspired by our previous research collaborations), let us consider a fictional data scientist named Ava who, along with non-technical clinical collaborator Ben, is building a model to predict whether a patient will be readmitted to a hospital in the future.
Existing literature and the landscape of available tools~\cite{santos_visus_2019,amershi_modeltracker_2015,zgraggen_squeries_2015} suggest that even without the need to incorporate domain experts' feedback, specifying a model for this task could be a challenging technical problem. 
In contrast to other types of models, building algorithms to predict future outcomes depends on making sense of \textit{temporal event data}, which can be sparse and irregularly spaced and therefore demand extensive wrangling and pre-processing~\cite{Gellad2023,Vaithianathan2019}.
For example, Ava might need to aggregate past diagnoses at varying time intervals and carry forward previous values to produce a clinically meaningful value at each hospital admission, which itself is an irregularly-spaced observation.
However, existing tools for data scientists largely focus on the more general, technical problems of ``data cleaning''~\cite{epperson_dead_2023,kandel_wrangler_2011} and model selection~\cite{krause_infuse_2014,amershi_modeltracker_2015} without considering the temporal nature of the data.
As a result, a representative workflow for Ava's task might consist of building an elaborate feature extraction pipeline in SQL, performing modeling using Python in a Jupyter Notebook, then exporting model summary statistics to a slide deck to share with Ben. 
If during their discussion of the model's behavior Ben identified a specification issue in the model, Ava would likely have to rewrite sizable portions of the code, then execute the rest of the analysis over again.

Human-AI interaction literature reflects the ad-hoc nature of model specification and the difficulties of bringing non-technical domain experts into the loop.
Most studies with decision-makers often take place only \textit{after} a model has been fully developed and/or deployed~\cite{Kawakami2022partnerships,Kuo2023,Zytek2021,Beede2020,Grgic-Hlaca2018,Sivaraman2023}.
This delays opportunities for experts to identify critical issues until after a team has largely committed to a specification, such that changes to the specification represent a significant step back in progress.
For example, a model that alerted clinicians to patients with sepsis was found to be largely redundant with clinical judgment during deployment~\cite{guidi_clinician_2015}, leading to a subsequent overhaul of the system focusing on longer-term prediction~\cite{Ginestra2019}.

Ethnographic work with data science and domain expert teams has highlighted the importance of collaboration in finding these pitfalls, but the processes depicted in these studies are unstructured and time-consuming because of the technical challenges described above~\cite{van_den_broek_when_2021,passi_problem_2019}.
Accelerating this iterative process may require tools to not just help domain experts visually inspect models after the fact~\cite{amershi_modeltracker_2015,cabrera_zeno_2023}, but also to support the more technical process of converting a modeling idea into a working prototype. 
Therefore, this paper seeks to address the following research question: \textit{How can we enable data scientists and domain experts to collaboratively specify and prototype models so they can be evaluated and improved on earlier in development?}

To answer this question, we introduce Tempo, an interactive system with novel technical affordances that can help data scientists and expert decision-makers work together more closely in the model specification process.
From prior literature describing collaborative efforts to design predictive models, we distill four design opportunities to support early predictive model development that are under-served by existing tools.
The resulting system shortens paths from ideation to evaluation by helping data scientists develop \textbf{readable yet computable queries on temporal event data} using a novel query language.
Tempo \textbf{generates model prototypes} based on the cohorts, input features, and target variables that the user specifies, enabling rapid exploration of many alternative specifications. 
The system also includes an interactive interface for \textbf{subgroup discovery}, which helps domain experts discover and probe rule-based data subsets with interesting characteristics. 
This enables users to assess and compare how models resulting from different specifications might behave in practice.
By reducing the complexity of model specification and prototyping while maintaining expressiveness, these design features aim to minimize the gap between programmer-facing representations and those that experts can easily evaluate.

We demonstrate Tempo's effectiveness in helping data scientists and domain experts collaboratively reason about predictive models through three case studies, including web browsing behavior analysis, decision support for critical care, and readmission prediction in home health care. 
Our results suggest that bringing experts into the loop earlier can spark productive discussions, more quickly reveal when a modeling goal is non-viable, and lead to promising new specification choices.
We also find that it can be beneficial to frame model specification as a dedicated opportunity for expert feedback during the machine learning development cycle, suggesting directions for future ML tooling.

We contribute:

\begin{enumerate}
    \item \textbf{An analysis of challenges in predictive model specification} that have arisen in prior human-centered evaluations and ethnographic studies of predictive modeling;
    \item \textbf{Tempo, an interactive system} that incorporates simple yet precise formats for specification, prototyping, and interpretation; and 
    \item \textbf{Three case studies} demonstrating how Tempo helps cross-disciplinary teams develop more relevant and useful model specifications.
\end{enumerate}

\section{Related Work}
\label{sec:related-work}

Our work is inspired and informed by challenges in predictive model specification that have been discussed in human-centered AI evaluations, reviewed in Sec. \ref{sec:related-spec-issues}.
To design and implement our system to address these issues, we expand on prior technical work around working with temporal event data (Sec. \ref{sec:related-event-data}) and interactively evaluating models (Sec. \ref{sec:related-subgroups}).

\subsection{Predictive Model Specification for AI Decision Support}
\label{sec:related-spec-issues}

Temporal predictive modeling represents a large and long-standing class of machine learning problems that aim to forecast a future action or outcome based on prior history.
Predictive modeling is currently employed in an extremely wide range of applications, including predicting sales~\cite{tsoumakas_survey_2019}, user intent~\cite{Caruccio2015}, global climate~\cite{Dueben2018}, and medical treatment~\cite{Sivaraman2023}, as well as more contentious topics like child welfare~\cite{Kawakami2022partnerships} and criminal justice~\cite{Grgic-Hlaca2018}.
However, issues of model specification have often been a major obstacle to adoption of predictive models, especially for decision support in high-stakes domains.
In the medical setting, model specification problems have been increasingly discussed in the literature as predictive modeling has grown more sophisticated and widespread; for instance, Sherman et al.~\cite{sherman_leveraging_2018} showed that calculating input features retrospectively around the time of known outcomes can cause models to perform misleadingly well in evaluation.
Unexpected behaviors may also arise because historical data is biased with respect to decisions that a DST is designed to influence. For example, in Caruana et al.'s pneumonia risk prediction model~\cite{caruana_intelligible_2015}, asthma appeared to lower mortality risk because it was associated with more aggressive treatment.
In sepsis, another promising disease area for DSTs, studies have pointed out the difficulty of choosing the right predictive target, given that both mortality and need for treatment may be imperfect signals for disease severity~\cite{Sivaraman2023,Sendak2020}.

Model specifications have also been discussed in the AI fairness literature as a lens to understand how DSTs may reflect or distort human decision-maker values.
For example, Kawakami et al.'s \cite{Kawakami2022partnerships} study with child maltreatment screeners highlighted concerns that their DST used public support use as a proxy for mental health and substance abuse concerns, thereby unfairly penalizing families that seek help. 
Tal~\cite{tal_target_2023} conceptualizes all predictive targets as inherently imperfect approximations, which can reflect nuanced, value-laden negotiations between data scientists and domain experts~\cite{passi_problem_2019}.
While explainable and transparent AI designs are often cited as a way to help decision-makers understand when the AI is misaligned with themselves in the moment~\cite{Zytek2021,Kawakami2022dis}, this often places additional cognitive burden on non-technical end users and reduces the DST's value~\cite{Sivaraman2023}.
Instead, we aimed to develop a tool that involves domain experts in model development from the start, ideally improving the alignment of the end product. 

\subsection{Querying, Visualizing, and Modeling Event Data}
\label{sec:related-event-data}

Temporal event data is ubiquitous as a way to represent evolving real-world processes and interactions through discrete observation. 
Storing and querying event data is a primary objective of many database systems, including OpenTSDB~\cite{opentsdb}, Timescale~\cite{timescale}, and Prometheus~\cite{prometheus}.
Query languages such as TSQL2~\cite{bohlen_how_2006}, the Event Query Language~\cite{wolf_introducing_2019}, and others~\cite{rozsnyai_sari-sql_2009,bry_high-level_2006} offer syntactic ways to apply temporal logic to event data, often following Allen's foundational framework of interval relations~\cite{allen_maintaining_1983}.
However, event queries still tend to be verbose and hard to read in commonly-used ML data wrangling tools such as the Structured Query Language (SQL), the Pandas library in Python, SAS, and the \texttt{dplyr} package in R. 
Moreover, even in systems specialized for event data, it can be challenging to write and update queries because different types of aggregations require dramatically different implementations (e.g., tumbling, hopping, sliding, and variable-length windows)~\cite{bohlen_how_2006}.

Considerable data visualization research has focused on helping analysts interpret temporal event data, particularly in healthcare contexts~\cite{guo_survey_2020}.
Systems such as DecisionFlow~\cite{gotz_decisionflow_2014}, Frequence~\cite{perer_frequence_2014}, and EventAction~\cite{du_eventaction_2016} aim to identify frequent event patterns and event sequences associated with adverse outcomes. 
Other systems such as VizPattern~\cite{jin_interactive_2010} and $(\text{s}|\text{qu})\text{eries}$~\cite{zgraggen_squeries_2015} have augmented the query-centric approach described above with visual representations.
While these systems tend to focus on exploratory analysis and temporal relationships between events (e.g. B after A), our work addresses the more ML-focused task of helping users aggregate events at standard intervals (e.g. A occurred four times in the past hour).

Despite the many approaches to \textit{visualize} event data, tools to address the challenges of working with event data for \textit{ML modeling} are comparatively scarce. 
For instance, systems by Kwon et al.~\cite{kwon_retainvis_2019} and Guo et al.~\cite{guo_visualizing_2019} help users interpret sequence models, but they assume the model specification and input data are fixed. 
Tempo addresses this gap by including model specification through temporal queries as a central part of the interface.

\subsection{Interactive Tools for ML Model Analysis}
\label{sec:related-subgroups}

Supporting ML practitioners in reasoning about and improving models has been a widely-studied problem in human-computer interaction and data visualization~\cite{Hohman2019}. 
Most closely related to our work are tools that help users navigate AutoML modeling results~\cite{ono2020pipelineprofiler,santos_visus_2019} and tools to evaluate and adjust data labels based on model behavior~\cite{amershi_modeltracker_2015,bhattacharya_exmos_2024,fiebrink_human_2011}. 
In particular, Visus~\cite{santos_visus_2019} and EXMOS~\cite{bhattacharya_exmos_2024} both allow users to configure the model specification to some degree before evaluation. 
Unlike these tools, however, Tempo directly supports event data, making it applicable to many temporal prediction problems where it is otherwise challenging to define specifications.

Recently, model interpretation and evaluation tools have increasingly adopted \textit{subgroup-level} (sometimes also called slice-based or rule-based) analyses of model behavior. 
Subgroup analysis originates from data mining, in which classification and frequent itemset algorithms are applied to find data subgroups with interesting distributions of a given metric~\cite{herrera_overview_2011}.
Several slicing tools are aimed toward ML model analysis, mining data slices with higher error rates than average~\cite{chung_automated_2020,eyuboglu_domino_2022}.
Visual analytics tools such as SliceTeller~\cite{zhang_sliceteller_2022}, Zeno~\cite{cabrera_zeno_2023}, and others~\cite{kahng_visual_2016,zhang_manifold_2019} have utilized discovered and manually-curated subgroups to help data scientists characterize model errors.
Our system extends these workflows by allowing the user to define custom metrics within and across model specifications, interactively edit and evaluate rules, and visualize subgroup feature values across timesteps.

Few existing ML tools have as their primary goal the development and refinement of model specifications. Amershi et al.~\cite{amershi_examining_2010} explore how helping end-user ML creators understand how target variables and labeling schemes may or may not support their needs, but their system is tailored towards small-scale concept learning rather than predictive modeling at scale. Cashman et al.~\cite{cashman_user-based_2019} introduce the concept of exploratory model analysis, similar to Tempo's workflow but without the feature extraction and subgroup evaluation steps. Finally, Dingen et al.'s RegressionExplorer system~\cite{dingen_regressionexplorer_2019} allows clinicians to evaluate model specifications on non-temporal data using handcrafted subgroups. These systems demonstrate the potential for interactive model specification tools to improve the quality and applicability of resultant models; our work expands upon their ideas to address the unique challenges of temporal models and decision support.

\section{Challenges in Predictive Model Specification}
\label{sec:challenges-design-opps}

Data science work often centers on the inherent tension between the real world and how to model it. As Passi and Jackson describe it, data scientists must ``continuously straddle the competing demands of formal abstraction and empirical contingency'', applying their discretion and prior experience to find abstractions that will yield useful results~\cite{passi_trust_2018}. 
To achieve usefulness, data scientists must not just translate broader objectives into abstract modeling problems, but also negotiate these translations with non-data-scientist stakeholders~\cite{passi_problem_2019}. 
To better understand how technical systems could support this process, we review and analyze in detail four case studies in which the authors document how predictive model specifications have evolved throughout their projects.
To our knowledge these are among the only papers that directly address the development of predictive modeling specifications, perhaps because it is not often explored in simulated study settings and because there is little incentive to publish nonviable specifications.
Nevertheless, these four studies, drawn from communities ranging from anthropology to HCI to medicine, put data science practices in dialogue with downstream stakeholder perspectives and complicate the traditional linear narrative from identified needs to models, informing the design of our system. 

\paragraph{Case 1: Proxy outcomes for special financing.}
\citeauthor{passi_problem_2019}~\cite{passi_problem_2019} present an ethnographic account of data science and product teams at a large company working on making data-driven predictions about customers who need car financing. 
In developing a predictive model to match potential borrowers with auto dealers, the teams updated their understanding of what constituted a good match several times.
These evolutions occurred in response to both data issues raised by the model developers (e.g., high rates of missingness in important variables) and insights shared by business analysts in response to initial results (e.g., alternative data sources to try).
At each iteration of the problem specification, teams interrogated the accuracy of the model against the relevance of its outputs, showing that even with the most promising outcome variable, accuracy was \textit{``at best, slightly better than a coin flip.''}
The project ultimately failed to produce a viable model, which one business analyst attributed to general challenges with their population of interest: \textit{``There are literally thousands of reasons [for bad credit] that aren't that capture-able.''}
Determining the right problem specification thus required lengthy, nuanced negotiation between the business objectives and the realities of the available data.

\paragraph{Case 2: Identifying ``good'' hiring candidates.}
In a two-year ethnographic field study, \citeauthor{van_den_broek_when_2021}~\cite{van_den_broek_when_2021} show how data scientists, human resources (HR) experts, and managers navigated the implementation of a new algorithmic aid to screen job candidates.
 Although the team initially intended to favor algorithmically-mined insights over experts' presupposed notions of how to design the model, they ultimately leaned on expert knowledge throughout data selection, outcome variable formulation, and assessment of the results.
For example, when presenting the model to the HR team, they identified highly predictive variables that experts had not originally considered important, potentially suggesting new and useful insight.
However, the experts also reasoned that if the model excluded features they considered important for the \textit{future} workforce they envisioned, it might simply reproduce undesirable \textit{past} hiring patterns.
Yet even after updating the model to include expert-selected features and deploying it, the team found that the hiring managers tasked with using the model rarely followed its suggestions, implying that \textit{``the algorithm was unable to provide useful insights into candidates.''} 
Without clear ways for hiring managers to see how the algorithm's outputs should influence their decisions, they found it difficult to align its predictions with their existing processes.

\paragraph{Case 3: Screening families for child maltreatment.}
\citeauthor{Kawakami2022partnerships}'s study with child welfare agency workers who review referrals for potential maltreatment~\cite{Kawakami2022partnerships} echoes the above concerns about outcome variable formulation and its impact on what types of cases an algorithm selects.
Screening workers in the study described being most focused on immediate risk to children, whereas the algorithm they used (the Allegheny Family Screening Tool, or AFST) aimed to predict longer-term risk using available administrative data.
Perhaps due to its reliance on prior records, instead of identifying families that were \textit{``slipping through the cracks''} of human assessment, workers saw the AFST as surfacing people who were \textit{``tripping on every single crack that they seem to encounter.''}
The algorithm was initially developed without significant input from the decision-makers, leading to surprise when workers encountered inexplicably high or low scores.
In such situations workers \textit{``made guesses about how each of these features might be influencing the AFST score,''} reasoning at the level of subgroups such as truancy cases or families with several referred individuals.
These practices allowed them to make use of the AFST in certain types of cases, while ignoring it when they saw the model's problem specification as more misaligned with their priorities.

\paragraph{Case 4: Clinician adoption of early warning alerts.}
Reports by \citeauthor{guidi_clinician_2015}~\cite{guidi_clinician_2015} and \citeauthor{Ginestra2019}~\cite{Ginestra2019} describe the development and prospective evaluation of two successive AI tools to detect severe sepsis (a life-threatening condition) in the hospital.
The first version of the tool was focused on \textit{detection} of patients with signs of sepsis~\cite{guidi_clinician_2015}, resulting in predictions that were largely seen as redundant because the model was predominantly finding patients who were already being monitored or treated for sepsis. 
To help identify cases that were not yet obvious to clinicians, the authors then developed a new early warning system that could predict \textit{future} onset of sepsis. 
Unfortunately, the new predictive version was less favorably received and resulted in fewer changes to patient management than the earlier detection model, perhaps because of \textit{``patients’ clinical stability at the time of alert''} and the \textit{``lack of established action items to implement after an alert.''}
Synthesizing the results of the two evaluations, the authors hypothesize that their models may have been seen as less useful despite their high performance because they were deployed and evaluated in overly broad populations.
They suggest that future work can target predictive tools to sub-populations that are both under-flagged by clinicians and most likely to benefit from early treatment~\cite{Ginestra2019}.

\subsection{Design Opportunities}
\label{sec:design-opps}

The case studies above illustrate that problem specification for predictive modeling is a highly ad hoc process that can often yield poor results even with significant time investment and opportunities for collaboration.
We propose four broad opportunities that interactive tools can address to structure this process and increase its likelihood of success:

\begin{enumerate}
    \item[\textbf{D1}] \textit{Balance technical flexibility with readability to enable both data scientists and non-data scientists to interrogate model specifications.} In all four cases, model specifications may have been somewhat opaque to non-data-scientist stakeholders because of the technical complexity required to implement them. Cases 1 and 2 describe data science teams curating one-off presentations to show stakeholders modeling results, making it difficult for these stakeholders to directly engage with model behavior. By including model specifications directly alongside interactive evaluations, tools can help data scientists and domain experts alike to examine the effect of specification choices. In order to support both types of expertise, however, such tools would need to provide both flexibility for data scientists to implement the algorithms they need, and readability for domain experts to interpret modeling choices. \label{goal:flexibility-readability}
    \item[\textbf{D2}] \textit{Cultivate more agile paths from model ideation to evaluation.} Cases 1 and 2 show that predictive modeling often requires long development time from establishing an initial formulation to obtaining prediction results.  Tools that make it easier to prototype models on complex temporal datasets, then update them in response to expert feedback, could help avoid unnecessary effort on refining models with less preferable specifications. \label{goal:accelerate-loop}
    \item[\textbf{D3}] \textit{Enable domain experts to interpret and critique model specifications in terms familiar to them.} Across cases 2, 3, and 4, domain experts and stakeholders identified ways that model specifications resulted in undesired behaviors.  However, these issues were generally only found after the team was largely committed to a particular model specification. Tools could help experts understand model behavior earlier in the process by mining and presenting interpretable patterns in their predictions and errors on validation data.\label{goal:interpret-and-critique}
    \item[\textbf{D4}] \textit{Support reflection on problem specification issues beyond target variable selection.} While some existing work encourages reflection and iteration on the target variable for prediction~\cite{guerdan_groundless_2023,jacobs_measurement_2021,amershi_examining_2010}, cases 2, 3 and 4 highlight that problem specification can also involve value-laden decisions around the input features and subpopulations used. These concerns can significantly impact the acceptability of predictive models but are currently difficult to systematically reason about. \label{goal:full-specification}
\end{enumerate}

\section{System Design}

We introduce Tempo, a visual analytics system that facilitates collaborative iteration on predictive model specifications. 
Tempo aims to help teams find potential pitfalls faster by speeding up and improving transparency in three major phases of early model development: expressing a model specification in terms of aggregations on temporal data, training an initial prototype model, and evaluating model predictions.
Below we first describe a usage scenario for how Tempo supports these tasks, then we detail the novel algorithmic and interactive aspects of Tempo's implementation.

\subsection{Usage Scenario}

\begin{figure*}
    \centering
    \includegraphics[width=\textwidth, alt={Two screenshots of Tempo, one showing the Models sidebar and the Specification Editor, and the other showing the Metrics view. The Models sidebar lists four models with different predictive targets, such as Readmission 30 Days and 90 Days. The Specification Editor shows input variables such as history of anemia, arrhythmias, arthritis, and others. The Target Variable shows a query for whether there is a future admission from 1 day in the future to 30 days in the future. The Metrics view shows a warning that the model could be approximated with fewer variables, including the time since the last admission. The model has an AUROC of 76\%.}]{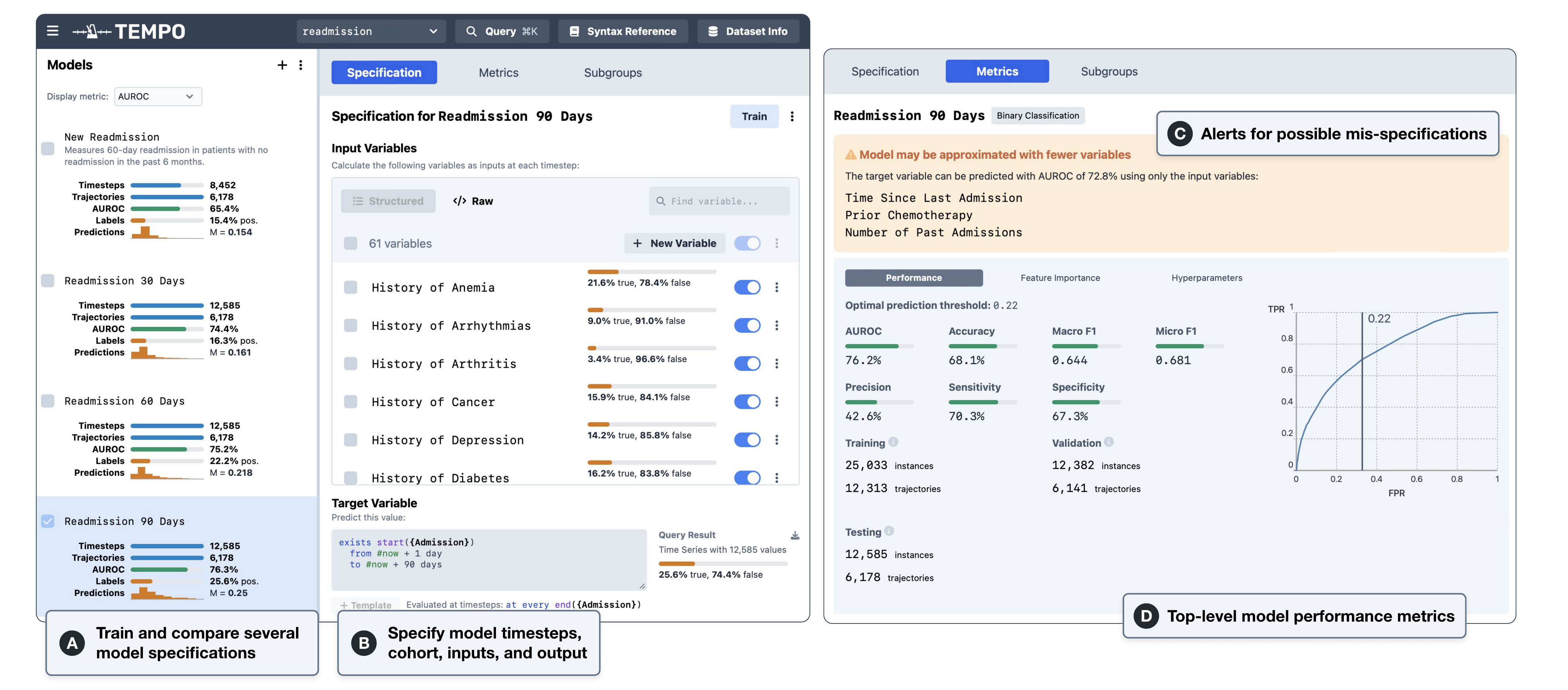}
    \caption{Model specification and training interfaces in Tempo, shown on an example use case of predicting readmission to a hospital from open-source clinical data with varying cohorts and time horizons. Users can create multiple model prototypes, listed in the Models sidebar (A), and edit how their inputs and outputs are defined in the Specification Editor (B). Once a prototype is trained, the Metrics view displays possible mis-specification alerts (C) and performance metrics (D).}
    \label{fig:spec-and-metrics}
\end{figure*}

Let us return to the scenario described in Sec. \ref{sec:introduction}, in which data scientist Ava is tasked with building a hospital readmission model.
Without Tempo, this task might require writing a large amount of code and making choices around what model behaviors to communicate, none of which would be easily visible to clinical collaborator Ben.
Using Tempo, Ava can transform their dataset for modeling and perform subgroup analysis interactively with Ben, helping them iterate on model specifications much more quickly.

Ava's first step is to import their data into Tempo.
Although some SQL might be required to download the data from a database, these queries would not require aggregation, making them much simpler to write than an entire pre-processing pipeline.
Ava writes simple queries to import dozens of different types of time-stamped events, such as hospital admissions, diagnoses, and prescriptions\footnote{In examples and screenshots throughout this section, we utilize MIMIC-IV data~\cite{johnson2020mimic} for about 6,000 patients.}.
These events can be irregularly spaced in time and are not guaranteed to align with admissions, which would make them challenging to work with in other languages.
After importing into Tempo, Ava can use the system's query language to aggregate the diagnoses and procedures at the end times of every admission, effectively creating a table with one row for every admission that contains the patient's clinical history.

\begin{figure*}
        \centering
        \includegraphics[width=\textwidth, alt={Screenshot of the Subgroups view in Tempo with callouts for interface elements, applied to finding groups with high positive true label rates in the Readmission 90 Days model. The top returned subgroup is patients with more than 20 past admissions. A callout for the Search Criteria button shows the editing interface for defining how subgroups are ranked. Another callout shows that when clicking on a subgroup, the interface shows an interactive list of Distinguishing Features for the subgroup compared to the overall dataset.}]{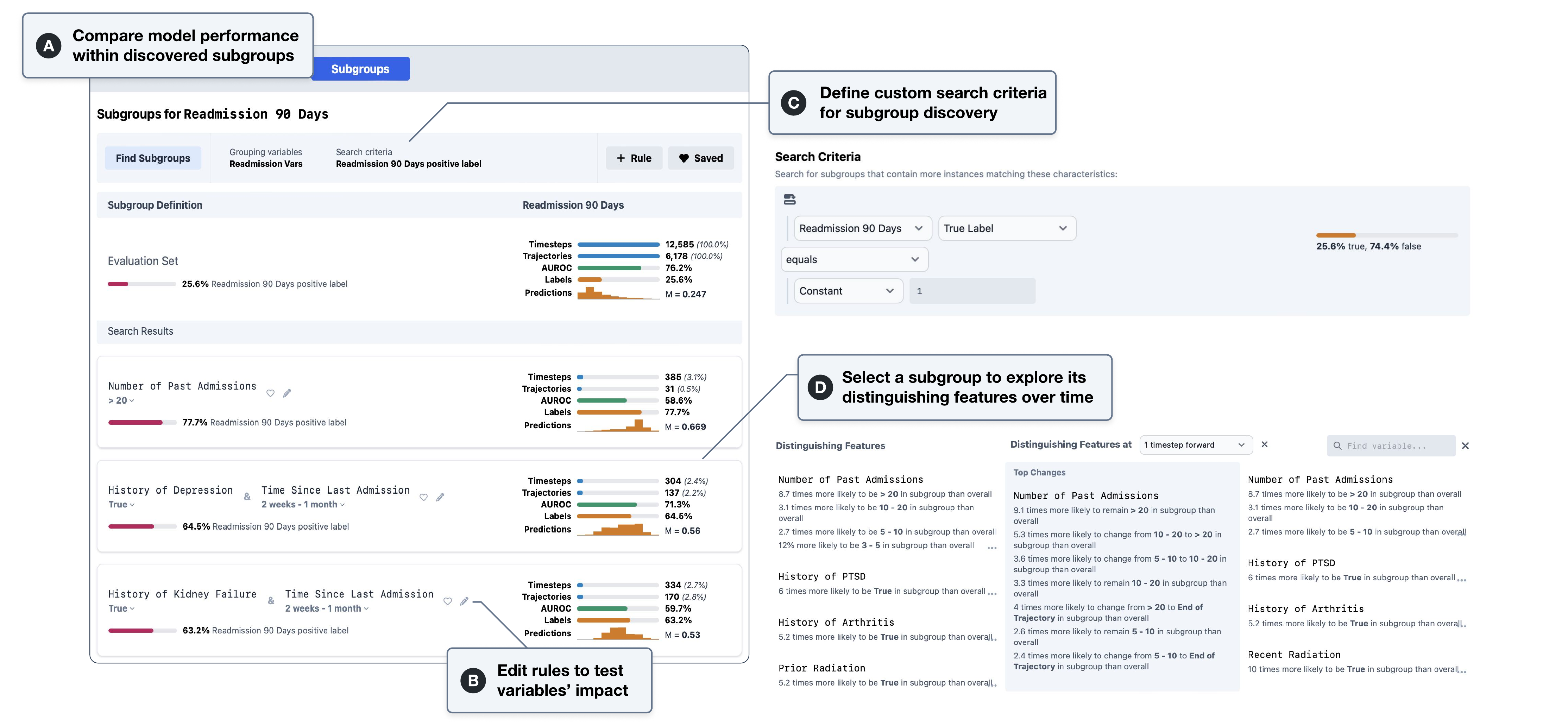}
        \caption{Subgroup discovery interface in Tempo. The Subgroups tab (A) allows data scientists and domain experts to automatically mine rule-based subsets of the data that have interesting characteristics. Users can edit and refine subgroup definitions to evaluate the impact of individual variables or try new combinations (B). Search criteria can be flexibly defined within and across models (C); here we show subgroups with positive labels for the 90-day readmission outcome specified in Fig. \ref{fig:spec-and-metrics}. Selecting a subgroup reveals the Distinguishing Features view, which lists additional variables that differentiate the given subgroup from the overall dataset over time. For instance, in the subgroup of patients with a history of depression and a prior admission between 2 weeks and 1 month ago, patients also tend to have many more past admissions and a history of post-traumatic stress disorder (PTSD).}
        \label{fig:slices-view}
\end{figure*}

Using the Specification Editor (shown in Fig. \ref{fig:spec-and-metrics}), Ava fills in the Timestep Definition, Input Variables, and Target Variable to create a model predicting readmission at 30 days. 
However, since the time horizon for readmission could lead to important differences in model behavior, Ava uses the duplication tool to quickly prototype two additional models measuring readmission at 60 and 90 days.
Whereas such a task would previously have required managing multiple dataset variants and increasing the complexity of their code, the ease of editing these variants in the Specification Editor allows Ava to efficiently explore more diverse modeling options.

Tempo automatically trains XGBoost models~\cite{chen_xgboost_2016} for each specification, resulting in top-level performance measures displayed in the Metrics view. 
Ava and their clinical collaborator Ben might look at the Model Metrics view for their 90-day readmission model (shown in Fig. \ref{fig:spec-and-metrics}C), and notice that it has overall good performance, with an AUROC of 76\%. 
However, an alert informs them that comparable accuracy can be attained using only three variables, two of which relate to past admissions and one represents prior use of chemotherapy. 
Ben reasons that past admissions could be a reasonable predictor of future re-admission, but that clinicians would likely already know the patient was high-risk based on their past visits. 

Ava and Ben can further investigate the model's performance in the Subgroups view (Fig. \ref{fig:slices-view}), which allows them to search for subsets of the data with interesting characteristics.
They notice that the top subgroup with a high rate of positive true labels consists of patients with over 20 past admissions, who are over 3 times as likely to be readmitted than an average patient. 
This observation adds weight to their hypothesis that most of the patients surfaced by this model specification would be due to prior admissions, suggesting that a clinical alert based on this model would be largely redundant with clinicians' knowledge. 
They then decide to explore other model specification ideas that could hone in on patients with a greater potential value-add for prediction, such as those without a recent admission.
Compared to the static slide deck they would have used, Tempo's interactive subgroup analysis tools help Ava and Ben reach mutually understandable conclusions faster.

\subsection{Temporal Query Language}

At the core of what makes Tempo useful to data scientists is its novel query language, which simplifies common but complicated pre-processing tasks while also providing greater transparency for domain experts.
A technical description of the syntax and how queries are computed can be found in Appendix \ref{app:query-syntax}; here we describe at a high level how Tempo helps write and verify temporal queries.

\paragraph{Importing Data into Tempo.}
Tempo's data format requires minimal analysis choices, so users can import as much data as possible and refine it later using queries.
The system supports three basic data types: \textit{Attributes} (data fields that remain constant throughout a trajectory), \textit{Events} (observations from a single moment in time), and \textit{Intervals} (observations that span a period of time with a start and end point). 
For example, in a medical setting, Attributes could include demographic features such as date of birth, ethnicity, and gender; Events could represent diagnoses and drug prescriptions; and the Intervals table could encode periods where a patient was admitted to the hospital. 
Any of these data types can be omitted if not applicable for a particular dataset.

\paragraph{Temporal Aggregations.}
The primary functionality of Tempo's query language is temporal aggregation.
Many temporal modeling tasks involve aggregating data at a fixed series of timesteps, sometimes called an index date: for example, counting all diagnoses for a patient within the last 30 days of each visit, or averaging the prices of the products a user has purchased prior to viewing a new item in an e-commerce setting.
However, it can be challenging in existing query languages to deal with unevenly-spaced timesteps and sparse or missing events, leading to less readable code.
In Tempo's syntax, aggregations take a similar form regardless of spacing and missingness considerations: \tqlinline{<aggregation function> <data field> <aggregation bounds> <timestep definition>}.
For instance, for the diagnosis example above, we could use a \tqlinline{count} aggregation function on a \tqlinline{\{Diagnosis\}} data field, specify the past 30 days as the aggregation bounds, and every visit for the timestep definition.
This would result in the following query:

\begin{lstlisting}[style=tempoql]
count {Diagnosis} from #now - 30 days to #now 
  at every {Visit}
\end{lstlisting}

Tempo also includes special syntax for common pre-processing routines to help create useful features for predictive modeling (addressing \ref{goal:flexibility-readability}). 
For example, the \tqlinline{impute} command can be used to replace missing entries with a fixed value. 
Conversely, the \tqlinline{where} keyword can be used to introduce missing values while preserving time series alignment. 
To discretize numerical features, the \tqlinline{cut} command allows users to define a binning strategy using custom or automatic cut points. 

\paragraph{Comparison with Existing Query Languages.}
\label{sec:simplicity-comparison}
\begin{figure*}
    \centering
    \includegraphics[width=\textwidth, alt={Two example temporal queries illustrated using a diagram and implemented in Tempo, SQL, and Pandas/Python. The first query checks whether there exists a heart failure diagnosis in the past 30 days at 30-day intervals. The second query checks whether the diagnosis exists anytime in the past, evaluated at the start of each hospital admission. The Tempo queries are each three lines long; the SQL queries are 12-17 lines; and the Python code is 14-17 lines.}]{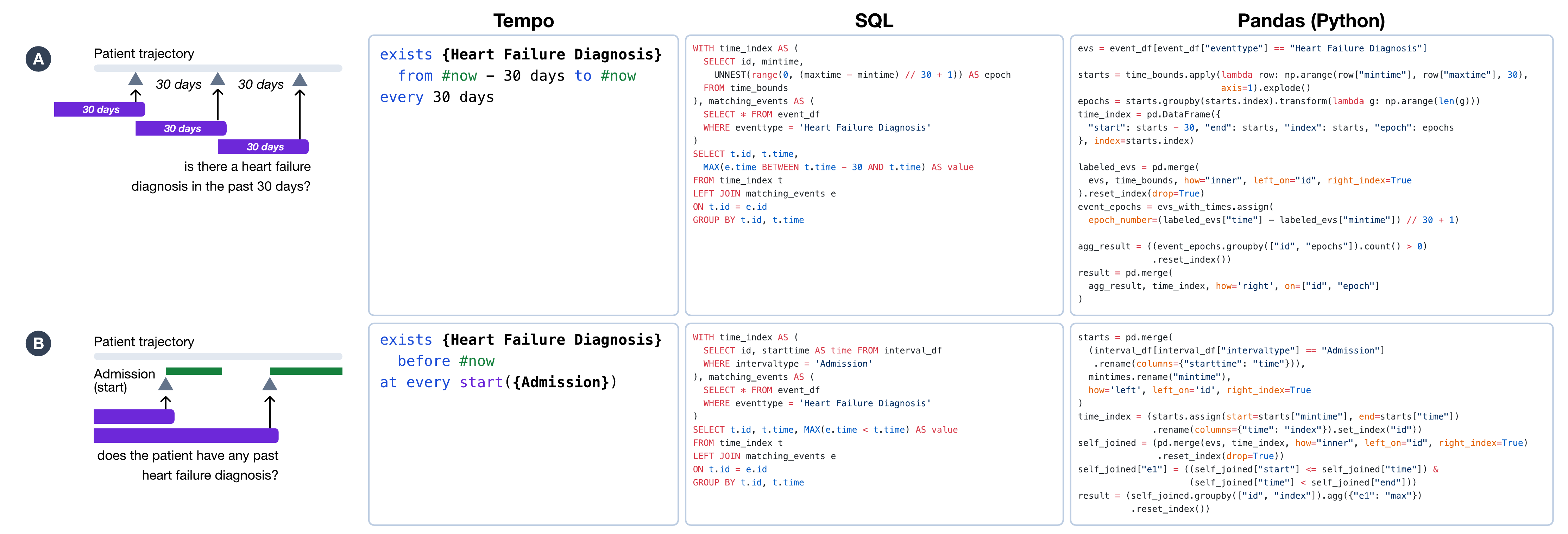}
    \caption{Two aggregations that might be performed on patient trajectories and how they could be implemented in Tempo, SQL, and Pandas. The Tempo queries are considerably more succinct, and do not require the user to keep track of multiple joining tables as SQL and Pandas typically do for temporal aggregations. Also, the differences between the two Tempo examples are small and semantically meaningful because no code optimization is needed to handle large datasets. (These examples all assume data is structured in Tempo's input format, which would resemble a typical database structure for this type of query).}
    \label{fig:code-comparison}
\end{figure*}
In contrast to the tools most commonly used by data scientists to pre-process model features (such as SQL, Pandas, and SAS), Tempo is explicitly designed to make temporal aggregations simpler to write and easier to read, even for non-data scientists.
To illustrate this comparison directly, Fig. \ref{fig:code-comparison} shows two plausible aggregations that a data scientist might need to implement for a predictive model: checking for an occurrence of a heart failure diagnosis every 30 days (Fig. \ref{fig:code-comparison}A) and finding any prior occurrence of the diagnosis at the start of a hospital admission (Fig. \ref{fig:code-comparison}B).
In Tempo, the two queries are roughly the same length as it would take to describe them in plain English, and their differences are directly related to the conceptual differences in the period and frequency of aggregation.
On the other hand, SQL and Pandas require highly verbose, multi-step operations to produce a correct result, and a substantially different approach is needed for the two queries. 

\paragraph{Supporting Query Authoring and Understanding.}
\label{sec:query-result-tiles}
To make working with Tempo's query language easier, the editing interface includes syntax highlighting and autocomplete to help users verify that a query is written correctly.
Users can reference the Dataset Info and the Syntax Reference page to learn what fields and functions are available.
Additionally, users can select from common functionality in the Templates dropdown menu, and fill in the resulting placeholders with their data to write queries.

\begin{figure}
    \centering
    \includegraphics[width=\linewidth, alt={Three example Tempo queries shown alongside Query Result tiles, which show compact histograms and bar charts describing the summary statistics for the query result. The first query checks for medical procedures containing the word "ventilation", which returns an Events object with several distinct string values. The second calculates the mean heart in the past 4 hours at every 4 hours, and the result is a continuous-valued Time Series with 69.3\% missingness, highlighted in red. The third query checks if there is any Diagnosis event containing the string heart failure in the past 90 days, and imputes 0 if the value is missing, at every start of a hospital admission. The result is a Time Series that contains "true" 4\% of the time.}]{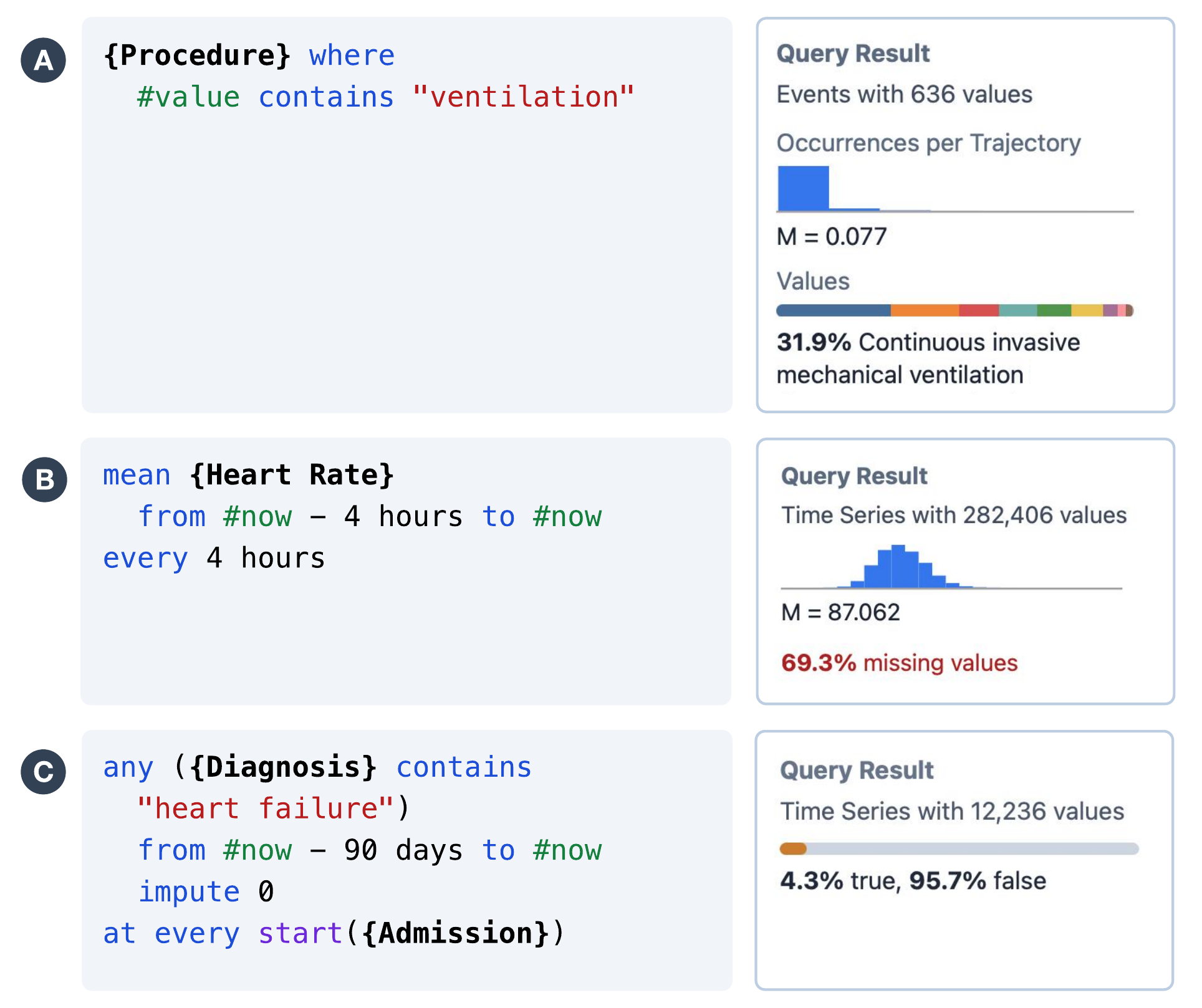}
    \caption{Examples of Query Result tiles (right) displayed when the user executes each of the queries (left).}
    \label{fig:query-results}
\end{figure}

To help data analysts verify that their queries are correct, Tempo incorporates \textit{Query Result tiles} (shown in Fig. \ref{fig:query-results}) throughout the interface that automatically display summaries of the query outputs as the user types. 
The design of these tiles follows prior systems for data profiling~\cite{epperson_dead_2023}, which have utilized compact and live-updating data summaries to accelerate experimentation. 
In Tempo, each tile indicates the result's type and length, the distribution of its frequency and duration if applicable, and the distribution of its values. 
Distributions are shown using histograms for continuous data and stacked horizontal bars for binary and categorical data. 
When the query output contains missing values, the missingness rate is shown in red to call the analyst's attention to this potential data issue. 

\subsection{Defining and Training Models}
\label{sec:model-definitions}

Together, the features of the Tempo query language described above can enable data scientists to develop a complete and succinct description of the data used to develop an ML model, addressing \ref{goal:full-specification}. 
In Tempo's Specification Editor (Fig. \ref{fig:spec-and-metrics}B), four sections determine how a model's inputs and outputs will be defined:

\begin{enumerate}
    \item \textit{Timestep Definition.} As with the aggregation expressions described above, the model's timestep definition specifies at which time points the model will be run.
    \item \textit{Timestep Filter.} Optionally, times selected by the timestep definition can be removed if they do not match a predicate query. This can be used to define inclusion or exclusion criteria for entire trajectories or for individual timesteps.
    \item \textit{Input Variables.} The bulk of the model specification, the Input Variables consist of a query for each feature to be used as input to the model. Each query is evaluated against the model's Timestep Definition to ensure that the resulting Time Series are aligned. If a query returns a categorical variable, it is automatically converted to a one-hot encoding.
    \item \textit{Target Variable.} This field specifies what to attempt to predict. Like Input Variables, the target is evaluated against the model's Timestep Definition; timesteps with a missing value are excluded from model training.
\end{enumerate}

By default, all models use an \mbox{XGBoost} model architecture~\cite{chen_xgboost_2016}, which is widely used across predictive modeling domains and enables users to quickly get evaluation results with good performance and generalizability (addressing \ref{goal:accelerate-loop}). 
Users can also choose to use a variety of neural network architectures to implement the prediction, including dense feedforward networks, recurrent neural networks (RNNs), or transformers.
The system performs a hyperparameter search  using Ray Tune\footnote{\url{https://docs.ray.io/en/latest/tune/index.html}}, then returns the model with the best performance on the validation set.
Tempo can also export the input and output matrices for training, validation, and test cohorts in CSV format, allowing data scientists to apply fully custom modeling techniques in their programming environment.

After specifying and training the model, an overview of its results appears in the Models sidebar (Fig. \ref{fig:spec-and-metrics}A), which displays the model's statistics in sparkline-style charts.
(Each ``model'' here refers to the combination of a specification over the input data and the trained model prototype.)
More detailed results are shown in the Model Metrics view (Fig. \ref{fig:spec-and-metrics}D), which shows standard top-level performance metrics for each model, such as AUROC, sensitivity, specificity, F1 score, and $R^2$ score for regression models. 
Consistent with existing model-building workflows, the metrics include SHAP feature importance scores for all features that are used to train the model~\cite{Lundberg2017}.
The Metrics view also displays alerts when model metrics may be misleading (Fig. \ref{fig:spec-and-metrics}C), suggesting when changes to the specification may be needed (\ref{goal:accelerate-loop}). Tempo currently supports two types of alerts:

\begin{itemize}[leftmargin=*]
    \item \textit{Trivial approximation alert.} A model may have a small number of input variables that can easily predict the target, either because these variables are deterministically associated with the target or because some timesteps need to be filtered out of the modeling cohort. To efficiently detect this situation, we train models that use only the top 5 input features ranked by XGBoost's feature importance metric. If these small models attain a performance  of at least 95\% of the full model's performance, the alert is raised.
    \item \textit{Rarely-predicted classes alert.} A common problem in multiclass classification problems is that the model fails to learn one or more rare classes. Tempo raises an alert when any class has a recall of less than 10\%, indicating that the target variable may need to be changed or a different classification threshold selected.
\end{itemize}

\subsection{Subgroup Analysis}
\label{sec:subgroup-analysis}

Addressing \ref{goal:interpret-and-critique}, Tempo supports \textit{interactive subgroup discovery} to help domain experts understand and interpret model specifications using the familiar concept of rule-based subpopulations.
Similar to how Tempo's query language provides a simple yet computable representation for cohorts, inputs, and target variables, we envision subgroup discovery as an intuitive yet precise tool for expressing model behaviors.

The goal of subgroup discovery is to find subsets of a dataset, each defined by the intersection of one or more discrete features, that exhibit meaningful differences from the overall data. 
Subgroup discovery has been extensively explored in databases and machine learning research~\cite{herrera_overview_2011,chung_automated_2020,zhang_sliceteller_2022}, suggesting it is a promising way to help domain experts make sense of model behavior. 
However, experts may care about several different metrics beyond simple errors, such as false positives or negatives or even comparisons across multiple models (i.e. instances misclassified by one model but correctly classified by another)~\cite{holstein_improving_2019}.
Existing tools typically perform exhaustive search and do not directly support different search criteria, limiting the amount of exploration that can be done on large datasets.

In the Subgroups view (Fig. \ref{fig:slices-view}), Tempo incorporates subgroup discovery using the Divisi algorithm~\cite{divisi}, which enables efficient, approximate, and configurable search for subgroups.
As with other subgroup discovery methods, Divisi requires that the input features be discrete or categorical. 
These variables are automatically generated from the input feature queries used for modeling, and they can be edited and refined using Tempo query language syntax if desired.
The discovered subgroups are ranked according to a combination of \textit{ranking functions} designed to surface interesting subgroups, including the rate of the outcome of interest, the size of the subgroup, and the number of features used to define the rule (more details in \cite{divisi}).

\paragraph{Mitigating False Discoveries.}
Performing too many subgroup analyses comes with the risk of false discoveries, i.e., subgroups that appear different than average due to chance alone. 
We mitigate this chance by splitting the input data into \textit{discovery} and \textit{evaluation} sets~\cite{green_subgroup_2021}, such that the initial scoring and ranking occurs on a different subset of the data than that displayed in the interface.
This provides preliminary assurance that an identified subgroup is at least robust enough to be different than average in both the discovery and evaluation sets.

\paragraph{Interactively Editing and Defining Subgroups.} Upon finding a subgroup of interest, users can probe that subgroup further using several tools (Fig. \ref{fig:slices-view}B). Clicking an individual rule feature temporarily removes that feature from the rule defining the subgroup, so users can assess how much that feature contributes to the subgroup's size and performance. Users can also open a dropdown for each grouping feature to alter the values that are defined to be included in the subgroup. To perform more significant edits, a lightweight editing dialog can be used to redefine the predicate using Tempo query language syntax. Finally, users can click the \textbf{+ Rule} button to define a fully custom subgroup rule. Together, these interactions can help experts experiment with other feature combinations to better understand model behavior patterns.

\paragraph{Comparing Subgroups Across Models and Timesteps.} The Subgroups view also helps users understand how subgroups differ from the overall dataset over time. Clicking on a subgroup reveals the Distinguishing Features view (Fig. \ref{fig:slices-view}D), which shows the features \textit{not} in the subgroup's defining rule that tend to be most different within the subset. 
To track changes in the trajectories within the subgroup over time, users can select a timestep offset and compare how values have changed relative to the timestep that matched the subgroup.

\subsection{Implementation Details}

Tempo is a web application built using a Svelte frontend\footnote{\url{https://svelte.dev}} and a Flask backend server\footnote{\url{https://flask.palletsprojects.com/}}. 
Users can launch Tempo locally on their own datasets by writing a lightweight JSON configuration file that points to the Attributes, Events, and Intervals files in CSV or Apache Feather format.
This configuration file also allows data scientists to specify the sizes of the training, validation, and test sets, which are automatically generated upon opening the dataset for the first time.
The system is open-source to enable wider use.\footnote{\url{https://github.com/cmudig/tempo}}

Internally, the Tempo query language is parsed using Lark\footnote{\url{https://www.lark-parser.org}}, and query computations are performed using Pandas with heavy use of Numba\footnote{\url{https://numba.readthedocs.io/en/stable/}} to accelerate aggregations. 
Query results for individual variables are cached as columnar Apache Feather files so that they can be quickly and separately loaded from disk for model training and subgroup discovery. 
Tempo's queries could be translated to SQL queries in the future, allowing users to more easily work with large databases.

\section{Case Studies}

To explore how Tempo could support the development of predictive model specifications, we conducted three collaborative model-building efforts in different domains. 
These case studies were conducted in an exploratory manner and followed typical real-world data science practices.
Through the resulting models and qualitative user feedback, we aimed to assess whether Tempo addressed our Design Opportunities (Sec. \ref{sec:design-opps}) and what new opportunities it posed.

In the first two case studies, we worked with the teams asynchronously to understand and address their goals, built initial models in Tempo, then engaged participants from the two teams in semi-structured interview and feedback sessions. 
During these sessions we gave participants access to Tempo with the initial models loaded, and we worked with them to perform a part of the model-building process that was relevant to their expertise. 
In the final case study, we collaborated with a group of researchers with expertise in pharmacy and biomedical informatics over a period of three months beginning after the first version of Tempo was developed.
The team members with data science expertise (included as co-authors on this paper) worked with the HCI research team to build an initial model prototype in Tempo, after which we elicited feedback from two additional pharmacy experts.

\subsection{Predicting Web Browsing Behavior}
\label{sec:case-study-web-browsing}

We worked with a start-up that was developing a web browser extension for tab management and had collected a large dataset of de-identified user browsing logs. 
They were interested in understanding whether the events they had captured could be used to develop predictive models to make proactive suggestions to users. 
To explore this question, we first extracted data for about 500 users during the first 72 hours that they used the browser extension, including Events for navigating to a URL domain (called \tqlinline{Navigation}, with the values hashed for privacy) and Intervals representing when each tab existed (\tqlinline{TabExists}) and when it was active on the user's screen (\tqlinline{TabActive}). 
This resulted in a total of around 432,000 Event rows and 180,000 Interval rows.

\paragraph{Gaining Intuition into Model Performance.} The first model we trained was to detect whether a user would return to a tab after opening it, a task the team envisioned using to surface tabs for later use or automatically remove old tabs.
We defined the timesteps as the times that any tab was activated onscreen: \tqlinline{at every start(\{TabActive\})}.
Then, we designed 26 input features, summarizing the number of navigation events and tabs activated over varying windows, the presence of common URL domains, and the time spent within each tab and page. 

As there was no obvious choice of time horizon for the model's target variable, we decided to train two versions of the model: one predicting the existence of a matching tab activation event after at least 5 minutes, and one after at least 1 hour. 

We presented this model in a joint session with two team members, one with expertise in ML (referred to as U1) and the other in product management (U2). 
At first, U1 questioned why the AUROC of both models was only around 0.7, because they expected that \textit{``a very small amount of web pages make the majority of people's visits,''} and those web pages would be very predictable.
To test this hypothesis in real time, we trained a different model in which the timestep definition was every \tqlinline{Navigation} event, and the target was whether or not the user would return to the navigated URL domain again.
This model had an AUROC of 0.83, indicating that revisiting a particular URL domain was indeed easier to predict than revisiting a tab.

\paragraph{Balancing Time Horizon Trade-offs.} The Subgroups view revealed that in the 1-hour model, users were much more likely to return to a tab if they had previously activated that tab more than 50 times (53\% vs. 28\% positive labels, observed in 3.3\% of timesteps); however, this feature was not surfaced in the subgroups for the 5-minute model.
This led us to hypothesize that the 5-minute model was picking up on more short-term usage patterns, which U2 deemed less useful for the current product goals.
Instead, U2 suggested looking at longer-term patterns but within a particular limit:
\begin{quote}
\textit{``If you're gonna return to it in more than an hour, }but\textit{ less than some other threshold... [we could] say, `it's worth keeping this open as a tab in your tab bar... because you're gonna get back to it.' ''}
\end{quote}

To try this, we duplicated one of the models and modified the target variable to search for a return to the tab specifically between 1 and 5 hours in the future.
The results in the Model Sidebar showed that this new model had comparable accuracy to the original ones, and the number of positive examples was also lower than both original models.
We analyzed the subgroups for this model first for positive predictions and labels, then for negatives.
The former showed similar results to the original 1-hour model, but the accuracy on these subgroups tended to be lower.
For the subgroups ranked by negative labels and predictions, the top rule was \texttt{Num Times Domain Viewed = 1 AND Num Unique Tabs Recent > 5 AND Time Since Creation < 5 mins}.
U1 interpreted this as representing instances where \textit{``this person has been clicking around quite a bit and opens this new domain now, and it's not very likely they will come back to it.''}
Combining these two subgroup analyses, we could conclude that there were intuitive cases in which a person would clearly \textit{not} return to a tab, but it would be harder to predict with certainty that a person \textit{would} return.

\paragraph{Reflection.} 
In this case study, Tempo helped us quickly run modeling experiments over a large dataset in response to non-data-scientist feedback (towards our design opportunity \ref{goal:accelerate-loop}).
It also allowed U2 to design a time horizon for the target variable that would capture the types of instances they found most relevant.
Beyond the models created during this session, U1 saw potential for Tempo to help with managing their other temporal analyses due to the simplicity of the query language (\ref{goal:flexibility-readability}).
They also pointed out the value of subgroup analysis to provide value to end-users even when the model's overall accuracy is low: \textit{``If we can attach some semantics or intuitive explanations to [subgroups], then we can use that in the product.''}
Subgroup discovery could therefore serve not only to help debug model specifications, but also to guide where algorithms could most feasibly be deployed.

\subsection{Sepsis Treatment in Intensive Care}
\label{sec:case-study-sepsis}

We collaborated with four experts (U3--U6) from a large hospital system to develop predictive models for sepsis.
Sepsis occurs when the body's response to an infection causes further harm, leading to organ failure and sometimes death~\cite{CentersforDiseaseControlandPrevention2021}. 
While AI has been developed to recommend treatments for patients with sepsis, prior work has shown that these systems tend to make misleading recommendations for patients with severe illness~\cite{Sivaraman2023}, indicating that better model specifications may be needed. 
To examine how best to predict patients' future need for treatment, we extracted a subset of roughly 14,000 patients with sepsis from MIMIC-IV, a publicly-available dataset of intensive care patient records~\cite{johnson2020mimic}.
The data imported into Tempo contained about 19.2M events of 89 different types (vital signs, lab values, physical assessments, etc.), and 1.2M intervals of 387 types (primarily drug administrations and procedures). 
We used these events to define about 300 input variables using Tempo’s query language, each aggregated over 4 hours prior to each timestep.

\paragraph{Excluding Uninformative Patients.} We showed U3 and U4, both experts in medical data analysis, an initial model to predict whether in the next 8 hours a patient would receive vasopressors, a class of drugs used as a second-line sepsis treatment.
In the Subgroups view, U3 observed that the groups with most positive predictions always included a variable for prior vasopressor use. U3 thus concluded that the model was not clinically valuable, as future predicted vasopressor use was highly contingent on prior treatment that clinicians would already be aware of. 
Accordingly, they used the Specification Editor to build a new model using a Timestep Filter to exclude patients that had received vasopressors in the preceding 8 hours.
U4 further built on this model to include only timesteps in which vasopressors had \textit{never} been used, yielding slightly higher accuracy.
As both their roles frequently involved extracting features for ML modeling, U3 and U4 noted that Tempo's low-code model specifications could help them perform feature selection alongside a clinician:
\begin{quote}
    \textit{``I could imagine sitting and being like, `Oh, hey! Here we just check-mark [some features]'... [or say] `I don't want to carry forward [missing values].' [The clinician] can in real time be seeing it, which is a little different than just... [me] making all the decisions.'' (U3)}
\end{quote}

\paragraph{Interpreting Model Errors.} We presented U5, a statistician experienced in ML modeling, with the model that U4 had built.
Comparing Tempo to their current workflows, U5 noted Tempo’s convenience in quick data manipulation: \textit{``It would otherwise take me quite some time to actually develop this model. And if I want to change anything, it's going to involve readjusting the data sets and readjusting the parameters.''}
They also extensively used the Subgroups view to understand the model's behavior, finding that the rules were \textit{``much better than a variable importance plot, where we usually just try to look at how the features are related to the prediction values... This is like trying to do phenotypes.''}
In particular, they wanted to investigate points of disagreement between actual outcomes and model predicted values (e.g., vasopressor was administered, but the model posited it was not going to be). 
They searched for subgroups with high rates of model error, hypothesizing that these could either represent cases where \textit{``these patients need more attention,''} or cases where the treatment was not as important to the patient's outcome. 

\paragraph{Evaluating Clinical Relevance.} Finally, we showed U6, a clinician, the same model to gauge how meaningful its predictions were. 
While overall the results in the Subgroups view matched their clinical intuition, they expressed surprise with some of the identified rules--for example, on a subgroup that included patients with low calcium, they noted that \textit{``there's some physiologic basis for low calcium leading to [septic] shock... but I don't typically use those.''} 
They were more skeptical of other features returned further down the list: \textit{``I don't think you could convince anyone to start [vaso]pressors just based on having a gout history.''}
Therefore, U6 suggested modifying the grouping variables in Tempo to explain the model's behavior using clinically intuitive features, even if the model itself was using whichever variables yielded the best accuracy.

\paragraph{Reflection.} 
This case study highlighted how subgroup analysis can help both data scientists and domain experts work with model specifications.
While the data analyst and statistician participants used the Subgroups view to find possible mis-specifications and evaluate model errors, the clinician participant (U6) used it to understand how the model worked and brainstorm how it could be presented to end users (\ref{goal:interpret-and-critique}).
Thanks to the model understanding they gained, U6 was able to brainstorm other specifications that went beyond refining the target variable (\ref{goal:full-specification}), such as assessing the patient when important clinical indicators changed.

\subsection{Readmission for Home Health Patients}
\label{sec:case-study-home-health}

For the final case study, we worked with an academic team of pharmacy and biomedical informatics researchers interested in making predictions for patients in home health care.
As the US population ages, the number of people with chronic diseases and other age-related maladies will increase, posing a great challenge for the healthcare system and financial support for health~\cite{landers_future_2016}. 
To cater to the preferences of older people and reduce the cost of healthcare delivery, US health insurance policies have begun to favor home health care (HHC)~\cite{donnelly_id_2016}. 
Home health care offers medical, therapeutic, and supportive services in the patient's home/residence, aiming to manage health, maximize independence, and alleviate the consequences of disability or terminal illness~\cite{jones_characteristics_2012}.
Studies have shown mixed results on the effectiveness of HHC in reducing future hospital readmissions~\cite{naylor_transitional_2004,bahr_nurse_2020}, suggesting that 

its utility might vary for different patient populations.
Therefore, predicting readmission for home health patients could help clinicians decide when to recommend HHC and help flag at-risk patients to HHC providers.

To begin designing a predictive model for readmission in home health patients, we obtained de-identified electronic health records from a large regional hospital system over a period from 2015 to 2018. 
From this large sample, we created a Tempo dataset containing the 7,537 patients who had both a prior inpatient visit (e.g., an emergency department visit or an overnight hospital stay) \textit{and} a period of time spent receiving home health care.
For each of these patients we extracted demographic information as Attributes, procedures and home health assessments as Events, and medical visits, diagnosed conditions, and prescribed drugs as Intervals.
This resulted in 2.2M Events and 18.7M Intervals, an average of 2.7k observations per patient.
An important factor in the data import process was deciding how to code the conditions and procedures, since medical vocabularies (SNOMED, ICD-10, CPT-4, etc.) vary in complexity and in their intended use cases.
We chose to leave our options open by including two separate interval types for conditions, one coded in a higher-level categorization and another using industry-standard SNOMED codes.

\begin{figure*}
    \centering
    \includegraphics[width=0.8\textwidth, alt={Two timeline diagrams for different model specifications. The first shows that the included population is patients with >1 in-patient or emergency room visit, and >1 home health care visit, and the exclusion criteria are under 18 years old. After the 180-day baseline collection period and the hospital discharge, there is a 30-day window for the target event, hospital admission or ED visit. In the second timeline, diagnosis of chronic kidney disease is added to the inclusion criteria, and an additional 3-day period to check for home health or skilled nursing admissions is added.}]{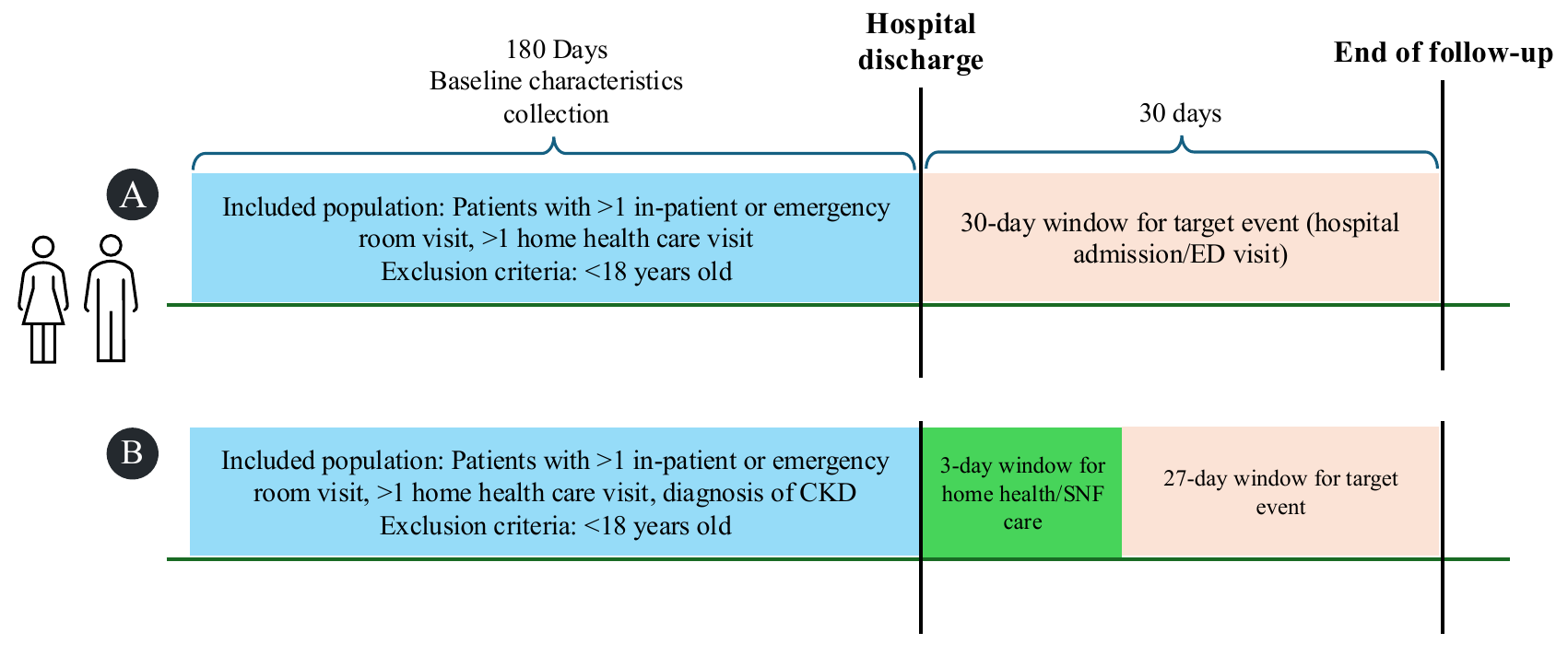}
    \caption{Two proposed model specifications for predicting readmission in home health patients. In both cases, hospital discharge events are used as the timestep definition, input variables are gathered from the prior 180 days, and the target variable of readmission is defined over the following 30 days. Based on domain expert feedback on specification (A), we add a 3-day window in (B) to check for admission to home health or a skilled nursing facility (SNF).}
    \label{fig:homehealth-diagram}
\end{figure*}

\paragraph{Building, and Ruling Out, ``Unsurprising'' Models.} Using prior knowledge about readmission in home health patients, the data science team developed an initial sketch of a model specification to implement in Tempo (shown in Fig. \ref{fig:homehealth-diagram}A). 
The model would be evaluated at the end of every in-patient visit, incorporate about 40 input variables, and aim to predict if the patient would be re-admitted at least two days after and within 30 days of discharge.
To get an initial sense of the feasibility of the model, our pharmacy data scientist collaborator learned to use the query language to build an initial specification.
They were able to learn to write temporal queries within a single session, finding Tempo's autocomplete and Dataset Info features particularly helpful in getting started.

We first implemented the specification in Tempo using only demographics and medical history (23 out of the 40 planned variables).
This yielded a AUROC of 58.5\%, fairly low but greater than random. 
We then showed this model to U7, a pharmacy expert with some experience in predictive modeling, and U8, a clinical pharmacist, in feedback meetings that lasted roughly 30 minutes each.
Both experts agreed that 30-day readmission was a useful target to predict from the perspective of lowering hospital costs: \textit{``Everybody tries to avoid a 30-day hospital readmission''} (U7).
However, upon seeing the trivial approximation alert listing history of heart failure, renal disease, and cancer as sufficient to predict readmission,
U7 emphasized that a model basing its predictions on these comorbidities would be completely unsurprising.
Instead, they suggested exploring factors that could help \textit{prevent} readmission within these diseases: \textit{``Was that a readmission for an electrolyte abnormality? Were people adherent [to their prescriptions]? That would be preventable.''}

The team therefore considered an approach where we would not predict readmission on all home health patients in a single model, but rather aim to find the driving factors of readmission within specific diseases.
To test out this approach in one such subpopulation in which our team had expertise, we created a new model in which the Timestep Filter was limited to patients with renal disease diagnoses.
However, this resulted in a very poor AUROC of only 51\%, likely because only 1,849 patients matched the inclusion criteria.
In other words, although the renal failure model's predictive specification would have been perceived as more useful, prototyping with Tempo quickly showed us that the lack of data availability would make this model largely infeasible.

\paragraph{Envisioning Other Modeling Goals.} Based on these less-than-promising results, the team ideated other ways to provide useful insights about the relationship between readmission and HHC.
For example, after examining the Subgroups view and seeing \texttt{Days in Hospital Last 6 Months} as a driver of predictions (see Fig. \ref{fig:teaser}), U7 suggested identifying \textit{``people who} shouldn't \textit{have gone to home health care.''} They verbally sketched a profile of a person who might do well with home health care as opposed to staying in a skilled nursing facility: \textit{``There's a 60-something-year-old, cognitively intact individual who looks like they can care for themselves. That person probably has less readmissions to begin with.''}
In contrast, an individual that had required more frequent care in the past might benefit from skilled nursing to avoid another readmission.

While this task was initially framed to support clinicians' decisions, U8 wondered how such a model would help them make discharge decisions given the constraints they usually face:
\begin{quote}
\textit{``Is it that I'm gonna intervene on this person and do some sort of special education, or, I don't know, something before they go home to try to keep them at home with home health care? Or is it more you're trying to show that home health care is not the right discharge disposition for the patient? And I think that gets harder because a lot of that is dictated by sort of what is going to be paid for [by insurance].''}
\end{quote}
Instead, U8 suggested that the model could be framed as a tool to \textit{``make a better argument''} to insurers about the discharge decision.
This led us to create the updated specification shown in Fig. \ref{fig:homehealth-diagram}B, in which we would split the data by the discharge location of either home health or skilled nursing care in the first three days after discharge.
We could then predict readmission using the discharge location as an input variable and see for which types of patients that feature had the most impact.

\paragraph{Reflection.} 
Through this case study, we found that Tempo's simple temporal aggregation features make it much easier to generate a model prototype on highly complex, messy data such as EHRs.
Unfortunately, while Tempo did accelerate the path from initial plan to functional model, it did not necessarily result in the team finding a viable specification.
However, by allowing us to gain insight into the predictability of readmission given the available data, we were able to quickly pivot away from less promising specifications and spend more time exploring alternatives (addressing \ref{goal:accelerate-loop}).

\section{Discussion}

We presented Tempo, an interactive tool that supports the process of specifying and evaluating predictive modeling tasks. 
Addressing design opportunities from prior work revealing how data scientists navigate model specification in practice, Tempo incorporates a novel query language to aggregate temporal event data and interactive subgroup discovery tools to probe model behavior in-depth. 
These features together addressed the primary research question in this work: how we can enable data scientists and domain experts to iterate on model specifications together and find potential pitfalls faster.
In our three case studies, the prototype models we created in Tempo were sufficient to surface model specification issues, in some cases revealing that the approach we were taking was very likely to fail. 
Tempo sparked discussion of these issues between data scientists and domain experts \textit{before} we had committed to any one approach, enabling us to quickly explore other, more promising directions.
Below we discuss implications and directions for future work based on the experiences of developing Tempo and using it in real-world contexts.

\textbf{Evaluating functional model prototypes surfaces diverse specification issues without the need for extensive model refinement.}
Because it is difficult to evaluate how AI will behave in practice while concurrently developing and improving AI capabilities~\cite{yang_re-examining_2020}, there are few opportunities during development for domain experts to determine whether a predictive problem is formulated appropriately.
Model development teams may bring assumptions on what technical approach will yield the most useful and accurate results~\cite{van_den_broek_when_2021,yildirim_sketching_2024}, and it is difficult to test these assumptions without substantial technical effort. 
In contrast, in our case studies the ability to create rough prototypes of models lowered the barrier for experts to critique their specifications and quickly resolve minor issues.
In cases where the expert feedback indicated that an entire approach was non-preferable (e.g. Sec. \ref{sec:case-study-home-health}), we were able to reuse and repurpose components of the model specification for new tasks, lowering the amount of effort required to shift the project's goals.

\textbf{Centering discussions about model specification leads to divergent exploration of the modeling space.}
Defining the predictive task in relation to its intended use case is often the first step in normative data science processes (e.g., the CRISP-DM model~\cite{chapman_crisp-dm_2000}).
However, despite its importance to the acceptability of the final model, the process of defining a predictive task is often ill-defined and requires ad hoc collaboration and iteration outside of existing tooling.
As a result, discussions about model specification may often be implicit, lack common language, or be embedded into more technical discussions about data pre-processing or model architecture selection.
Using Tempo, team members frequently posed speculative questions such as: 
What type of predictive target is most appropriate and fair for the decision-maker's task? 
At what time points should model predictions be computed? 
What variables should be used to explain its behavior? 
These discussions could be sparked even with very preliminary results as described in Sec. \ref{sec:case-study-home-health}, helping to generate ideas for experiments that may not have been obvious a priori but could lead to feasible and useful predictions.
Further evaluations are needed to evaluate whether this deeper exploration leads to more effective or acceptable models.
Nevertheless, the discussions Tempo facilitated are a first step toward expanding the space of possible specifications and avoiding concerns such as those raised around previously-deployed DSTs~\cite{Kawakami2022partnerships,passi_problem_2019,sherman_leveraging_2018}.

While the types of temporal predictions Tempo supports can be applied to a wide range of important use cases, our findings also suggest ways to cultivate divergent exploration of modeling tasks more generally.
For example, it could be valuable for analytics systems to surface how problem formulations--that is, for whom, what, and how predictions are made--affect behavior patterns in recommender systems~\cite{moller_designing_2024}, machine learning models for causal inference~\cite{dingen_roa_2024}, or generative models~\cite{korbak_pretraining_2023}.
Supporting these tasks may require new techniques to express specification choices (analogous to the Tempo query language) as well as to articulate and compare behaviors in domain-appropriate ways.

\textbf{When designing future specification tools, potentially with large language models (LLMs), both readability and precision are important for effective collaboration.}
Tempo adds to a growing body of literature in HCI examining how to make modeling choices visible and changeable to non-data-scientist experts~\cite{bhattacharya_exmos_2024,lam_model_2023,subramonyam_solving_2022,cabrera_zeno_2023}.
Feedback from our case study participants points to Tempo's potential to foster collaboration between data scientists and domain experts by providing value to people in both roles.
In particular, data scientists saw Tempo's query language as not only a way to iterate on model definitions alongside non-technical stakeholders, but also as a generally more convenient way to work with temporal data.
With the availability of large language models, future tools for model specification could lower the barrier to working with temporal data even further by helping data scientists generate and adjust SQL queries, or explain them to non-technical stakeholders. 
However, such solutions would still require a translation between implementation and explanation, creating the possibility of subtle LLM-induced mis-specifications or interpretation errors.
Future work should investigate how LLMs could support model specification in useful, robust ways, such as generating many variables according to a user-defined template or suggesting alternative specifications to try.
As opposed to simply generating modeling code from a natural-language description, these design approaches could reduce the friction of using a query language while maintaining its transparency. 

\textbf{Considering model specification as its own data science task results in a productive conceptual distinction from data extraction and cleaning.}
Many systems aim to assist data scientists ``clean data'' in preparation for modeling, a process that necessarily entails subjective choices with high impact on the final model performance~\cite{van_kuijk_preparing_2019,sherman_leveraging_2018}. 
Tempo supports many tasks on temporal data that might ordinarily be labeled data cleaning, such as variable definition, windowing and aggregation, and missing data imputation.
Because these choices can be made atop standard input data representations (Attributes, Events, and Intervals) and using Tempo's query language, they are more straightforward to implement, easier to validate, and more readable.
However, this has the effect of pushing some technical burden upstream of Tempo in the form of choices about how to format the data for import.
For example, in the home health case study (Sec. \ref{sec:case-study-home-health}), we updated the underlying Events and Intervals several times to add fields that had been omitted from the raw data, ensure that values across fields were coded consistently, and align data from several tables against a common set of trajectory IDs.
Although it may seem like an extra step to convert raw data into Tempo's format before being able to create the model inputs, this conceptual distinction helped separate choices that could be made largely for expediency (e.g., what coding scheme to use for prior conditions) and those that would require domain expert input (e.g., over what lookback period to aggregate the conditions).
Future data science tools could consider incorporating intermediate analytical representations of the raw data similar to Tempo's input data, especially if such representations could be generated automatically and refined later.

\textbf{Subgroup discovery is a promising approach to help domain experts interpret models at scale and at the point of decision-making.}
Whether and how to explain ML model outputs is an ongoing research area that typically requires a deep understanding of how the model should integrate into users' workflows~\cite{rong_towards_2024}.
For example, the child maltreatment risk screening tool described in Sec. \ref{sec:challenges-design-opps} was received with mistrust by users because of its lack of transparency, yet studies in healthcare have found that explanations can have minimal or even detrimental effects on decision quality~\cite{Bussone2015,Sivaraman2023}.
In contrast to feature explanations, which are one of the most common techniques used to help people assess the trustworthiness of model predictions~\cite{Gellad2023,Zytek2021,Sivaraman2023}, this work instead examined whether a rule discovery approach could better provide insight into model behavior at the subgroup level.
Rather than focusing on model errors or explaining instance-level predictions as in past work involving rule-based explanations~\cite{chung_automated_2020,zhang_sliceteller_2022,guidotti_2019}, Tempo enables users to search across different behavioral characteristics over an entire dataset at interactive speeds. 
Many experts in our case studies were already inclined to conceptualize data in terms of interpretable subgroups, such as patient phenotypes. As a result, they found the Subgroups view to be an intuitive way to reveal when certain types of trajectories should be excluded from training, when a model is most reliable, and even how to explain predictions within groups of similar instances.
Importantly, participants wanted to be able to edit and curate these subgroups according to the features they found meaningful.
Future work is needed to determine how best to support this curation process and translate the results into explanations that can be shown to end users.

\subsection{Limitations}

Because of the complexity in developing predictive models, the current version of Tempo only addresses a portion of the technical challenges involved.
Future versions of the system would likely need to directly interface with other data sources such as SQL databases and provide additional advanced editing capabilities for data scientists.
The query language could also be augmented to support more exploratory query types, helping analysts familiarize themselves with the data within Tempo.
Additionally, although Tempo supports training large deep learning models in addition to tree-based models like XGBoost, its focus is on enabling users to prototype models based on a specification rather than training the best possible model.
Without easy ways to edit many features at once, the system's reliance on human-crafted features may limit its usefulness for larger models. 
We note that for many applications in which predictive models are used, such as mixed-initiative user interfaces, medical screening tools, and financial services, lightweight and transparent models are preferable to deep neural networks. 
When a large model \textit{would} eventually be required, Tempo can help data scientists assess model formulations using preliminary neural network results and gain intuition about what can and cannot be predicted from their data.

Given the difficulty of quantifying success in any real-world model building effort, we chose case studies as an evaluation method to qualitatively explore the opportunities and potential challenges with collaborating around a model specification tool. 
In these case studies, although we worked with our data scientist participants to refine the initial models, most of the initial data cleaning and model prototyping work was done by us.
We also did not collect system logs of the actions users performed in Tempo, our focus being mainly on the models we explored and the users' feedback rather than the system's usability.
The feedback we received from data scientists suggests that it was largely intuitive to use Tempo's query language and subgroup analysis features.
Nevertheless, a longer-term, more controlled evaluation is needed to determine whether data scientists can easily learn to integrate a model specification tool like Tempo into their workflows.

\section{Conclusion}

If predictive models are to be accepted and used in important task domains, experts in those fields must find them to be relevant and useful additions to their decision-making processes.
While recent empirical studies have revealed challenges in identifying the right model specification, our work is among the first to explore how a technical system could address these challenges.
By making model specifications and behaviors salient and readable while still being precise and computable, tools such as Tempo can help bring non-data-scientist stakeholders into the loop earlier in development and surface ways that models might conflict with their intuitions. 
Identifying these misalignments earlier in turn opens a wider range of options to \textit{change} the model's behavior.
As predictive models become increasingly ubiquitous, reflecting on model specifications can help ensure that the final models are sensible to experts and beneficial to those their predictions may impact.

\begin{acks}
We are grateful to the data science and domain expert teams who worked with us to process their data for Tempo, used the system, and provided their feedback.
We also thank the following people for their helpful feedback on the concept and manuscript: Katelyn Morrison, Will Epperson, Dominik Moritz, Wesley Deng, Michelle Lam, Jill Lehman, John Minturn, Nur Yildirim, and John Zimmerman.
This work was supported by the Jewish Healthcare Foundation, by a National Science Foundation Graduate Research Fellowship (DGE2140739), and by the Carnegie Mellon University Center of Machine Learning and Health.
\end{acks}

\bibliographystyle{ACM-Reference-Format}
\bibliography{references}

\appendix

\section{Tempo Query Language}
\label{app:query-syntax}

Below we provide a more technical description about how Tempo's query language supports formatting data for modeling.

\subsection{Basic Queries} 
In Tempo's syntax, data fields are selected by wrapping them in curly braces, and they can be operated on using arithmetic and logical operations similar to SQL. By default, operations apply to the value associated with a data field (such as a temperature measurement), while the \tqlinline{time}, {\tqlinline{starttime}, and \tqlinline{endtime} functions can be used to extract the timestamps of the data field. For instance, to calculate the patient's age in years at the start of each admission, we could combine the timestamp of each \tqlinline{Admission} interval with a \tqlinline{Birth Time} attribute: 

\begin{lstlisting}[style=tempoql]
(starttime({Admission}) - {Birth Time}) as years
\end{lstlisting}

Attributes are automatically broadcasted such that each interval time is compared with the correct attribute value for that trajectory.

\begin{figure*}
    \centering
    \includegraphics[width=\textwidth, alt={Four representations of the same query. The first shows the desired output, which is every 4 hours from the patient's first to last observation, calculate the mean heart rate in the past hour. The second shows the Tempo query: mean Heart Rate from now minus 1 hour to now every 4 hours. The third shows the flow of computation, using a Timestep Definition to generate Aggregation Bounds which are each 1 hour long, then gathering Heart Rate Events that match those bounds and aggregating them together to produce the Aggregation Result. The final representation shows a diagram of the example patient trajectory, with aggregation intervals spaced 4 hours apart that last 1 hour each.}]{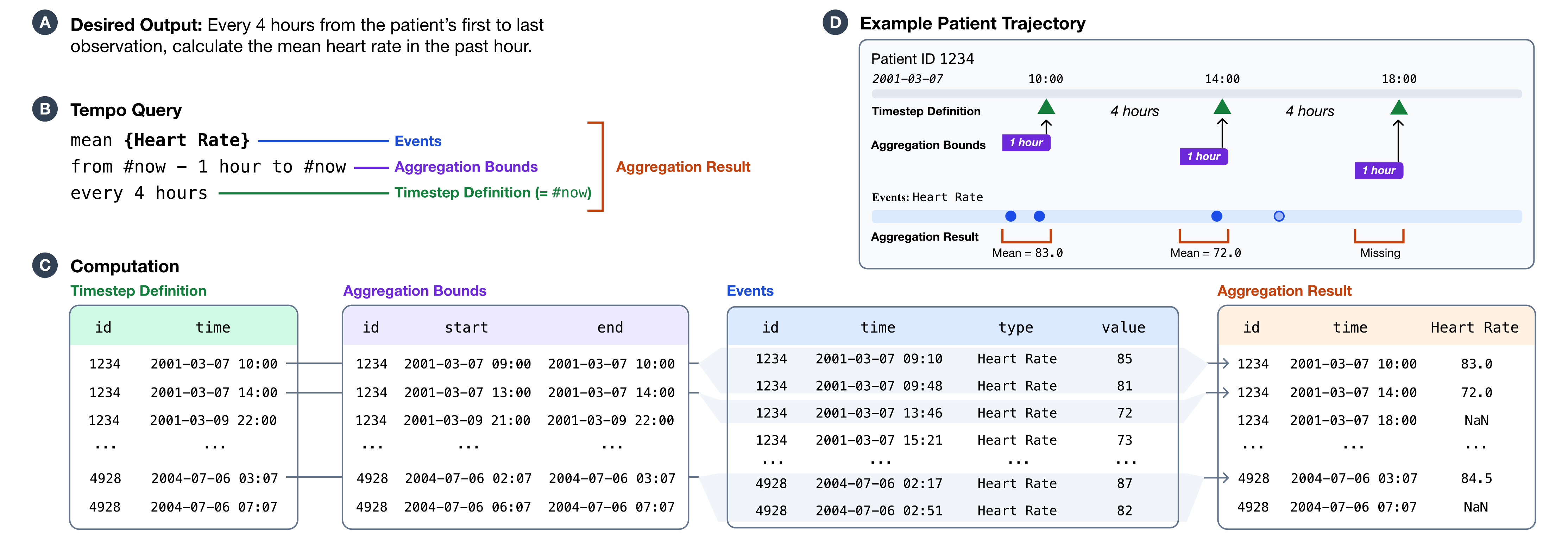}
    \caption{Example of using Tempo's query language to implement a standard temporal aggregation (A). The Tempo command for this query (B) is computed as follows (C): the Timestep Definition specifies at what times a row is included in the result, the Aggregation Bounds define over which intervals to search for Events in each row, and the Aggregation Result combines the matching Events by averaging them. As shown in the example diagram (D), this approach allows unevenly-spaced and potentially missing Events to be aligned with a standard set of patient IDs and times.}
    \label{fig:query-language}
\end{figure*}

\subsection{Aggregations} 
While basic operations on the raw data evaluate to Attributes, Events, or Intervals (which may be unevenly-spaced), aggregations result in a Time Series that is aligned to a user-defined \textit{timestep definition}, as shown in Fig. \ref{fig:query-language}. 
The timestep definition specifies the bounds and frequency of timesteps that should be selected within each trajectory. 
Each aggregation is computed over the observations that fall within the provided aggregation bounds, which are typically a function of \tqlinline{\#now} (which represents the time of the current timestep). 
Aggregation bounds make it equally straightforward to define overlapping, non-overlapping, and variable-length window sizes without concern for how the aggregation will be implemented. 
A variety of common functions are supported to produce the aggregation result, including \tqlinline{mean}, \tqlinline{min}/\tqlinline{max}, \tqlinline{any}/\tqlinline{all}, \tqlinline{first}/\tqlinline{last}, \tqlinline{exists}, \tqlinline{count distinct}, and others. 
For example, the query listed in Fig. \ref{fig:query-language}B calculates the mean of a patient's heart rate in the past hour, with one row for every four hours in the patient's trajectory.

Aggregations over Interval observations can be further customized by specifying whether the value of the observation should be treated as an amount or a rate. For example, if we know that an  \tqlinline{IV Fluid} Interval represents the total amount of intravenous fluids administered at a constant rate, we can use the  \tqlinline{sum amount} aggregation function to calculate the amount of fluid given between the aggregation bounds (even if these do not align with the interval bounds).

Note that Tempo aggregations automatically handle some subtleties that can render similar queries hard to implement correctly in other languages. For example, unlike conventional window functions, Tempo aggregations by default create a row for every timestep selected by the timestep definition, even if no value exists in the expression being aggregated. This makes it easier to align multiple Time Series to produce a matrix of feature values as long as they share the same timestep definition. Moreover, Tempo aggregations can be nested and the inner aggregation result will be automatically broadcasted, enabling aggregations that combine multiple non-aligned data fields.

\subsection{Convenience Syntax for Temporal Feature Engineering}

Because Tempo is explicitly designed to help create useful features for predictive modeling, it includes special syntax to make common pre-processing routines easier to read and modify. Combining these elements, we can perform several pre-processing operations within a single simple query.
For example, the below query could generate a Time Series containing quantile-binned classifications of each patient's average body mass index (BMI) using their most recent height and weight observations, imputing the median value when missing:

\begin{lstlisting}[style=tempoql]
(last {Weight} before #now) / 
  ((last {Height} before #now) ^ 2)
impute median
cut 3 quantiles named ["Low", "Average", "High"]
[timestep definition]
\end{lstlisting}

\end{document}